# Pinning-induced formation of vortex clusters and giant vortices in mesoscopic superconducting disks

W. Escoffier, I.V. Grigorieva
School of Physics and Astronomy, University of Manchester, Manchester M13 9PL, UK

V.R. Misko, B.J. Baelus, F.M. Peeters
Departement Fysica, Universiteit Antwerpen, Groenenborgerlaan 171, B-2020 Antwerpen, Belgium

S.V. Dubonos, L.Ya. Vinnikov
Institute of Solid State Physics, Russian Academy of Sciences, Chernogolovka 142432, Russia

ABSTRACT

Multi-quanta, or giant, vortices (GVs) are known to appear in very small superconductors near the superconducting transition due to strong confinement of magnetic flux. Here we present evidence for a new, pinning-related, mechanism for the formation of GVs. Using Bitter decoration to visualise vortices in small Nb disks, we show that confinement in combination with strong disorder causes individual vortices to merge into clusters or even GVs well below $T_c$ and $H_{c2}$, in contrast to well-defined shells of individual vortices found in the absence of pinning.

Mesoscopic superconductors, i.e., such that they can accommodate only a small number of vortices, are known to exhibit complex and unique vortex structures due to the competition between surface superconductivity and vortex-vortex interactions [see e.g. 1-6]. For mesoscopic disks, theoretical studies found two kinds of superconducting states: a giant vortex (GV), i.e., a circular symmetric state with a fixed value of angular momentum that can carry several flux quanta [1,2] and multivortex states (MVS) with an effective total angular momentum corresponding to the number of vortices in the disk (vorticity $L$) [3]. Recently, it became possible to experimentally distinguish between a single-core GV and a MVS composed of singly quantized vortices using the multiple-small-tunnel-junction method [4]. GVs have also been observed in rapidly rotating Bose-Einstein condensates (BEC) [7] and inferred from an experimental observation that vortices in perforated Al films merge into large flux spots very close to the superconducting transition temperature $T_c$ [8]. In another recent experiment, where Bitter decoration was used to directly visualise MVS in small Nb disks [5], circular symmetry was found to lead to the formation of concentric shells of vortices, similar to electron shells in atoms. Analysis of different vortex configurations revealed the rules ("periodic law") of shell filling and also "magic-number" configurations corresponding to commensurability between the shells (see e.g. [6]). Importantly, regular vortex shells could only be observed in disks with sufficiently weak pinning.

Here we show that, while weak pinning in mesoscopic disks only leads to distortions in symmetric shell configurations, the presence of strong disorder changes the situation dramatically. Using Bitter decoration to visualise vortices in small Nb disks and numerical simulations to analyse the role of disorder, we find that the interplay between confinement and pinning results in the formation of clusters or even GVs in relatively large samples and at temperatures well below $T_c$, due to *selective enhancement* of pinning strength by disk's boundaries. Depending on the disk's size and the applied field, we find that either all vortices merge into a single cluster/GV (typically for $L$=2 to 5-6) or clusters/GVs coexist with singly quantised vortices, for larger $L$. No clustering /GV formation was found for the same disorder in macroscopic samples.

Mesoscopic disks for this study were prepared from bulk Nb single crystals using e-beam lithography and reactive ion etching (CF$_4$) followed by high-vacuum annealing (at <10$^{-8}$ torr) at temperatures ~ 750 ºC. The above temperature is known to be sufficient to remove most contaminants, with reaction kinetics such that up to several days continuous annealing is necessary to achieve complete desorption [9]. Therefore, by varying the annealing time $t$ we were able to achieve partial removal of contaminants thereby changing the degree of disorder (in the form of non-superconducting inclusions) in the samples. The above produced large arrays of circular pillars - between 0.5 and 1.4 μm high, of three different diameters, 2, 3 and 5 μm - on the surface of ~0.5mm thick Nb crystals (see upper-right inset in Fig. 1a). The geometry of the arrays and the decoration details were the same as in our previous study [5], i.e., a whole array containing over 300 pillars was decorated in each experiment after field cooling to ~1.8K, allowing us to obtain simultaneous snapshots of up to a hundred vortex configurations in nominally identical pillars (further referred to as disks). In order to assess the degree of disorder before and after annealing, the samples were decorated immediately after etching, and then after 1, 2, 4 and 6 hours annealing at 750 ºC. No vortices were observed either in the bulk of the crystals or inside the disks for annealing times $t$ less than 2h, indicating that, as expected, contamination rendered the samples non-superconducting. Annealing for $t$= 2h or more recovered superconductivity over the entire crystals, with the amount of disorder gradually decreasing with increasing $t$ – see Fig. 1b,c. The disordered and very inhomogeneous vortex distribution in Fig. 1b - reminiscent of that observed in superconductors with high density of random columnar defects [10] - indicates the presence of many small pins. This is in agreement with previous studies, where dense arrays of 10-50nm normal inclusions were seen in Nb after a similar treatment [11]. Longer annealing times resulted in a much more homogeneous, although still disordered, vortex structure (Fig. 1c), very similar to that in macroscopic Nb films used in our previous study [5].

Fig. 1a (bottom-right inset) shows a typical vortex configuration in disks with weak disorder. Clearly, the effect of vortex confinement in single crystal-based disks is very similar to that in individual thin-film disks [5]. Indeed, previously confinement was found to have a mitigating effect on disorder in that, despite the presence of weak pinning, vortices formed nearly perfect shell configurations [5]; similarly, well-defined shell states are observed in weak-disorder disks on single crystals [e.g., state (1,6,13) in Fig. 1a]. Furthermore, both systems show identical diamagnetic response ($\Phi/\Phi_0$ vs $L$, where $\Phi=H \cdot S$ is the total flux through the disk's area $S$, $H$ the applied magnetic field and $\Phi_0$ the flux quantum), different shell states are observed over well defined intervals of $\Phi/\Phi_0$, etc. Details of the observed behavior will be published elsewhere. The only important implication for the present study is that the effect of vortex confinement in single crystal disks is the same as that of confinement in individual thin-film disks, even though in the former case only a small fraction of the total vortex length is subject to the additional interaction with the surface. Moreover, the response is independent of the disk's height, in agreement with theoretical predictions [12].

In contrast, in the case of strong disorder (as in Fig. 1b), instead of mitigating pinning as above, confinement was found to *selectively enhance* the effect of pins, ultimately leading to merger of vortices into clusters/GVs, as demonstrated below. The most striking features of vortex configurations in this case are, firstly, that some of the "vortices" inside the disks appear to be *significantly larger* than vortices in the bulk or in weak-disorder disks and, secondly, that there are on average noticeably *fewer* vortices in these disks, compared to the number expected for a given value of $\Phi/\Phi_0$ or found in weak-disorder samples, i.e., they show a stronger diamagnetic response and a significant variation in the observed value of $L$ for the same $\Phi/\Phi_0$, as demonstrated in the main panel of Fig. 1a. Indeed, a wide range of $L$ values is observed for any given $\Phi/\Phi_0$ (e.g. $L$=2-6, 8, and 9 for $\Phi/\Phi_0$=17-19 while only $L$=12 and 13 are found for the same flux interval in weak-disorder disks) and even the maximum observed $L$ is 25-30% lower than for the weak-pinning disks, indicating that strong disorder somehow facilitates expulsion of extra vortices. The two left insets of Fig. 1a show vortex images observed in the same experiment ($H$=70 Oe) in identical 3μm disks. Vortices in the bottom image form a slightly distorted shell configuration (1,7) while a disordered pattern of only 5 vortices is seen in the second disk of exactly the same area $S$, with one of the vortices having a ~2.5 times larger diameter compared to the rest. Indeed, identifiable shell configurations containing identical (small) vortices were seen only rarely in these samples while most disks contained a combination of small and large vortices in a disordered pattern or just a single large vortex. Fig. 2 shows typical observations for 2, 3 and 5 μm disks at $\Phi/\Phi_0$ corresponding to different maximum vorticities $L_{max}$ (i.e., maximum number of vortices found in disks with only small, singly-quantized, vortices). For $L_{max}$=3 most disks contained 2 or 3 vortices of a standard (small) size (top image), i.e., the diameter of a cluster of Fe particles 'decorating' a vortex was the same as for vortices in the bulk and in weak-disorder disks, but a significant proportion of disks showed only one vortex of about 80% larger diameter (bottom image). For $L_{max}$=9 only 10% of the disks showed configurations of small vortices (top image) while most of the rest contained a combination of several small and 1 or 2 large vortices (as in bottom image) and a few disks showed only one very large vortex, as in the bottom image for $L_{max}$=11. As the vorticity increased above $L_{max}$=11, very few disks contained only small vortices while a typical configuration was a combination of small vortices and a number of large vortices of several different diameters, as in the image for $L_{max}$=36.

To understand the nature of large vortices observed in disks with strong disorder, we recall that there is an excellent correlation between the size of vortex images in decoration experiments and the 'magnetic' size of individual vortices, $\propto \lambda$ [13], with only a weak dependence on vortex density. This allows us to identify the observed large vortices as several singly quantized vortices merged into GVs or clusters with vorticity $L^* \geq 2$. We emphasise that, although the presence of strong disorder appears to be the necessary condition for the formation of GVs/clusters, they are only formed in small disks and were never observed in the bulk of the same crystals. To estimate $L^*$ associated with different clusters/GVs we analysed intensity profiles of many different vortex images - see Fig. 3. Here we distinguish between the results obtained on disks containing only individual, only merged, or a combination of individual and merged vortices. One can see that (i) the size of individual vortices is well defined and shows only small variations between different disks and/or the bulk; (ii) merged vortices have only a few typical sizes (rather than a continuous distribution), with the same sizes found in disks with only one merged vortex (where the vorticity can be determined fairly unambiguously from $L$ vs $\Phi/\Phi_0$ curves) and in disks containing combinations of individual and merged vortices. This allowed us to identify the number of vortices merged in different clusters/GVs, $L^*$, as shown in Fig. 3. In the example of the disk in Fig. 3, this yields 5 individual vortices and one cluster/GV with vorticity $L^*$= 4; i.e., the total vorticity $L$=9, in very good agreement with $L$ expected for the corresponding value of flux through the disk and with the number of vortices observed in the same experiment in disks with only individual vortices. Similar agreement was found for our largest (5 μm) disks, such as the one in Fig. 2. According to the measured vortex sizes, here we observe 16 individual vortices with $L^*$=1, 3 merged vortices with $L^*$=2, 2 merged vortices with $L^*$=3 and one vortex with $L^*=5$. The total vorticity in the disk is then $L=33$, again in good agreement with the value of $\Phi/\Phi_0$ and the number of vortices found in the same experiment in disks containing only individual vortices.

To model the role of pinning in a confined geometry theoretically, we place a superconducting disk of thickness $d$ and radius $R$ in a perpendicular external field $H$. The forces of vortex interactions with each other $\mathbf{f}_i^{vv}$ and with the shielding currents and the edge $\mathbf{f}_i^s$ can then be modelled as [14,15]

$$\mathbf{f}_i^{vv} = f_0 \sum_{i,k}^{L} \left( \frac{\mathbf{r}_i - \mathbf{r}_k}{|\mathbf{r}_i - \mathbf{r}_k|} - r_k^2 \frac{r_k^2 \mathbf{r}_i - \mathbf{r}_k}{|r_k^2 \mathbf{r}_i - \mathbf{r}_k|} \right), \quad (1)$$

$$\mathbf{f}_i^s = f_0 \left( \frac{1}{1-r_i^2} - h \right) \mathbf{r}_i, \quad (2)$$

where $h = \pi R^2 \mu_0 H/\Phi_0 = (H/2H_{c2})(R/\xi)^2$, $\mathbf{r}_i = \boldsymbol{\rho}_i/R$ is the position of the $i$th vortex, $L$ the vorticity, and $f_0 = 4\pi\mu_0\xi^2 H_c^2/R$ the unit of force. Our numerical approach is based on the Langevin dynamics algorithm, where the time integration of the equations of motion is performed in the presence of a random thermal force. Then the overdamped equations of motion become: $\eta\mathbf{v}_i = \mathbf{f}_i = \mathbf{f}_i^{vv} + \mathbf{f}_i^{vp} + \mathbf{f}_i^T + \mathbf{f}_i^s$. Here $\mathbf{f}_i^{vp}$ is the force due to vortex-pin interactions (see, e.g., [16,17]) which is modelled by short-range parabolic potential wells located at positions $\mathbf{r}_k^{(p)}$; $\eta$ is the viscosity. The pinning force is

$$\mathbf{f}_i^{vp} = \sum_k^{N_p} \left( \frac{f_p}{r_p} \right) |\mathbf{r}_i - \mathbf{r}_k^{(p)}| \Theta\left(r_p - |\mathbf{r}_i - \mathbf{r}_k^{(p)}|\right) \mathbf{r}_{ik}^{(p)}, \quad (3)$$

where $N_p$ is the number of pinning sites, $f_p$ the maximum pinning force for each potential well, $r_p$ the pinning range, $\Theta$ the Heaviside step function, and $\mathbf{r}_{ik}^{(p)} = (\mathbf{r}_i - \mathbf{r}_k^{(p)})/|\mathbf{r}_i - \mathbf{r}_k^{(p)}|$. The $\mathbf{f}_i^T$ is the thermal stochastic force, obeying the fluctuation dissipation theorem $\langle \mathbf{f}_{\alpha,i}(t) \mathbf{f}_{\beta,j}(t') \rangle = 2\eta\delta_{\alpha,\beta}\delta_{i,j}\delta(t - t')k_B T$. The ground state of the system is obtained by simulating field-cooled experiments [18]. We now consider a disk containing, e.g., 8 vortices in the field $h = 15$ (for $R = 1.5\mu$m, this corresponds to $H \approx 45$Oe). For a perfect disk (no pinning), vortices form a symmetric two-shell configuration (1,7) (cf. [5]) as shown in Fig. 4(a). A weak pinning site placed near the center of the disk distorts the symmetric configuration, moving away the central vortex; the rest of the vortices adjust themselves accordingly (Fig. 4(b)). However, the situation changes dramatically if $f_p$ is increased: (i) a stronger pin traps 3 out of 8 vortices resulting in *vortex merger* into a cluster with $L = 3$ coexisting with individual vortices (Fig. 4c), similar to the experimental pattern shown in the inset of Fig. 3. A stronger still pin traps 5 out of 8 vortices (Fig. 3(d)). We note that the above pinning forces are of the same order of magnitude as those estimated for normal inclusions in Nb of radius $\geq\xi$ [19]. Furthermore, we find that the ability of a pin to trap vortices depends on its *position* inside a mesoscopic disk: a pin close to the disk's centre traps more vortices than the same pin near the boundary [see Fig. 4(e,f)], i.e., its pinning strength is *enhanced* when it is near the centre. (ii) The total vorticity $L$ in a disk with strong pins is *lower* than in the weak-pinning situation because vortices are repelled by a cluster stronger than by a single vortex, pushing them towards the disk boundary so that some vortices leave the sample. This explains the observed *enhanced diamagnetic response* of mesoscopic disks (see Fig. 1).

To clarify whether the above vortex merger corresponds to the formation of true GVs with a single core or multiquanta vortex clusters, we calculated the distributions of the order parameter, $|\Psi|^2$, and the phase, $\phi$, using the Ginzburg-Landau (GL) equations. It is known [3,4] that in perfect disks without pinning GVs appear as a ground state only if the disk's radius is small enough. For the disk sizes studied here, the ground state in the absence of pinning corresponds to vortex shells (MVS) as observed in ref. [5]. However, if vortices are trapped by a strong potential they can merge forming a GV because repulsive vortex-vortex interaction (logarithmic at small distances [20]) vanishes at very small distances $r_{min}$ when vortex cores strongly overlap [21]. To demonstrate the formation of GVs in disks with strong pinning, we introduced a pinning potential $U_{pin}(\rho)$ in the dimensionless GL equations (see, e.g., [15]),

$$(-i\nabla - \mathbf{A})^2 \Psi = \Psi\left(1 - U_{pin}(\rho) - |\Psi|^2\right), \quad (4)$$

where $U_{pin} = U_0 \exp(-\rho/w)$, and $\rho = \sqrt{\{(x-x_{displ})^2 - (y-y_{displ})^2\}}$. The results of calculations of $|\Psi|^2$ and $\phi$ are shown in Fig. 5(c,d) for a disk with $R = 20\xi$ and pins with $U_0 = -1$ and $w = 5\xi$ at the center and out-of-center, respectively. The Cooper pair density $|\Psi|^2$ vanishes at pin positions, forming large spots (larger than individual vortices): The central spot is a GV with vorticity $L = 5$, while the one at the distance of a half of disk's radius is a *cluster* consisting of $L = 3$ individual vortices. Thus, depending on the pinning strength, range and $r_{min}$ (see figure caption to Fig. 5), vortices can form either clusters (Fig. 5(a)) or GVs with $L = 5$ (center) and $L = 4$ (out-of-center), Fig. 5(b).

In conclusion, we present the first direct observation of pinning-induced merger of vortices into clusters or GVs in small superconducting disks with strong disorder. The interplay between confinement and disorder is shown to result in the *enhancement* of pinning by disk's boundaries so that up to 8 vortices can merge into a cluster or even a GV, with combinations of merged and individual vortices being most typical. The total vorticity in disks with GVs or clusters is reduced compared to disks without pinning, i.e., the system demonstrates an enhanced diamagnetic response. Our results can be generalized for other systems, e.g., vortices in BEC.

This work was supported by EPSRC, UK and the Flemish Science Foundation and the Belgian Science Policy. V.R.M. acknowledges support through POD.

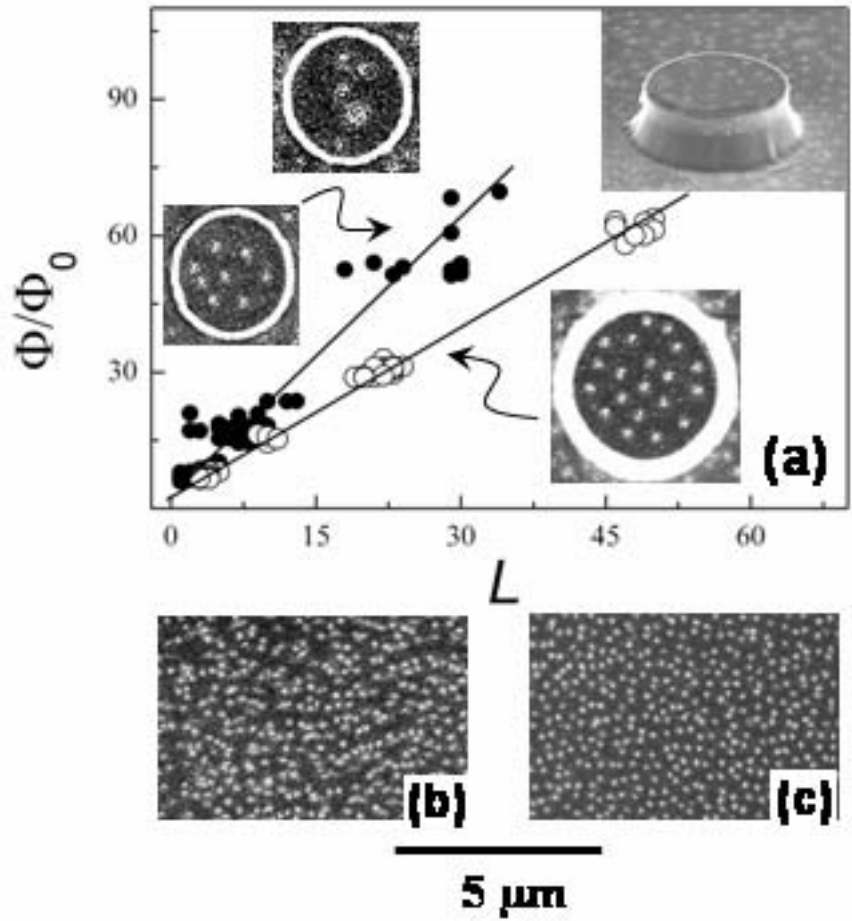

Figure 1. (a) Main panel: Diamagnetic response of disks with weak (○) and strong (●) disorder. Solid lines are guides to the eye. Top-left and bottom-right insets show vortex configurations in disks with strong and weak disorder, respectively ($H$=70 Oe). Top-right inset: vortices in and around a 1.4μm high disk decorated after field-cooling in $H$= 60 Oe. (b,c) Vortex distributions in the bulk (~100 μm away from mesoscopic disks) of Nb crystals with strong and weak disorder, respectively.

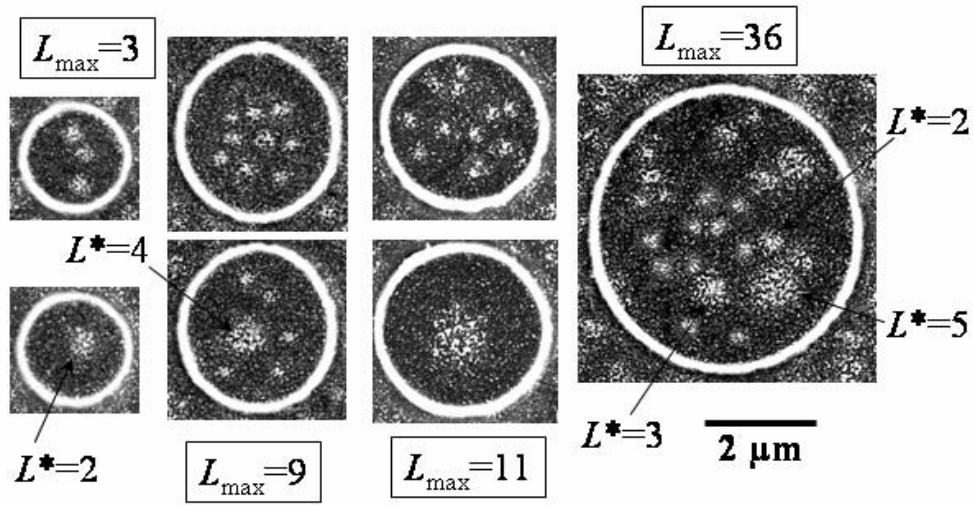

Figure 2. Typical vortex configurations in disks with strong disorder for different values of $L_{max}$ (see text). $L^*$ is the vorticity of GVs/clusters estimated from their diameters.

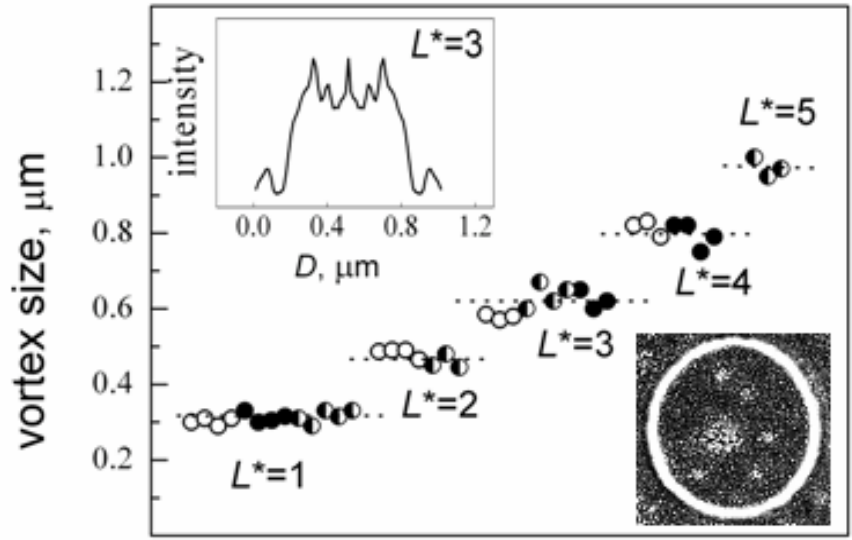

Figure 3. Diameters of individual and merged vortices determined from image intensity profiles on disks containing: only one vortex, either individual or merged (○); one merged and several individual vortices, as in the bottom-right inset (●); a mixture of individual and merged vortices of different sizes, as in the image of Fig. 2 for $L_{max}$=36 (◐).Top-left inset shows a typical radially averaged intensity profile for a vortex image. $L^*$ are estimated values of vorticity for clusters of different diameters.

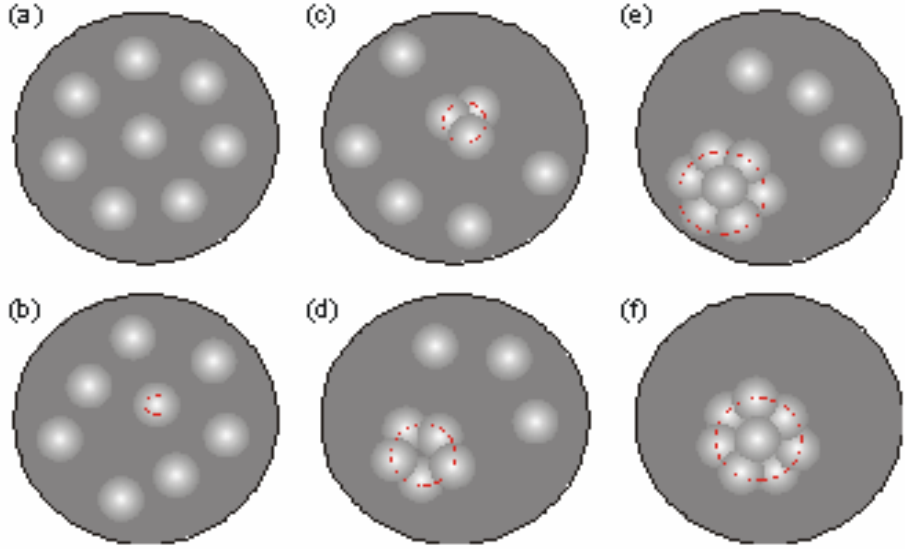

Figure 4. Vortex patterns for $L=8$ (a-d) and $L=10$ (e,f) in disks with different pinning potentials: (a) no pinning, perfect (1,7) state; (b) weak pinning, $f_p/f_0=2$, $r_p=0.08$, distorted (1,7) state; dashed line shows the range of pinning potential; (c,d) strong pinning, $f_p/f_0=4$, $r_p=0.16$ and $f_p/f_0=6$, $r_p=0.24$, respectively; combination of merged and individual vortices; (e,f) $f_p/f_0=8$, $r_p=0.32$: (e) pinning site close to the boundary; combination of merged vortices with $L=7$ and 3 individual vortices; (f) *enhancement* of the pinning strength for a pin near the center: $L=8$ vortices are merged, while other vortices are *expelled* from the disk by strong repulsion from the cluster.

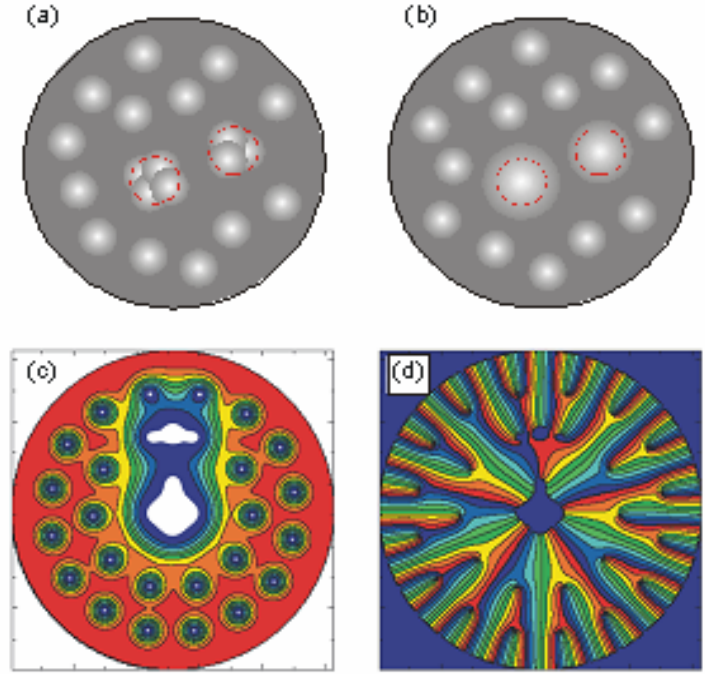

Figure 5. (a,b) Vortex patterns for $L=20$; formation of multiquanta vortices in disks with $f_p/f_0=10$, $r_p=0.16$ and different $r_{min}$: (a) $r_{min}=0.048$, vortices form clusters with $L=5$ (near center) and $L=4$ (out-of-center); (b) $r_{min}=0.064$, GVs with $L=5$ (near center) and $L=4$ (out-of-center); (c) Cooper pair density, $|\Psi|^2$, in a disk with $R=20\xi$, for $H=0.2H_{c2}$, for two pinning sites with $U_0=-1$, and $w=5\xi$, at the center and out-of-center; (d) phase pattern corresponding to (c) showing a GV with $L=5$ (center) and a cluster with $L=3$ (out-of-center).